\documentstyle[12pt]{article}

\begin{document}
\begin{titlepage}
\begin{flushright}
BA-95-21 \\
hep-ph/9506261\\
\end{flushright}
\begin{center}
\LARGE
{Proton Decay and Realistic Models of\\
Quark and Lepton Masses}\footnote{Supported in part
by Department of Energy Grant \#DE-FG02-91ER406267} \\
\vspace{.6cm}
\large
{\bf K.S. Babu}\footnote{Address starting September 1995: School of Natural
Sciences, Institute for Advanced Study, Olden Lane, Princeton, NJ 08540}
and {\bf S.M. Barr}\\
\normalsize
\vspace{.2 cm}
{\em Bartol Research Institute \\
University of Delaware \\
Newark, Delaware 19716 }
\vspace{.3 cm}
\end{center}
\def\strut{\rule[-.5cm]{0cm}{1cm}}
\def\sspace{\baselineskip = .18in}
\def\tspace{\baselineskip = .5in}

\begin{abstract}
\sspace
It is shown that in realistic SUSY GUT models of quark and lepton
masses both the proton decay rate and branching ratios differ
in general from those predicted in the minimal $SU(5)$ supersymmetric
model. The observation of proton decay, and in particular
the branching ratio $B[(p \rightarrow \pi^+ \overline{\nu})/(p
\rightarrow K^+ \overline{\nu})]$, would thus allow decisive
tests of these fermion mass schemes.  It is shown that the charged
lepton decay modes $p \rightarrow K^0 \mu^+, p \rightarrow K^0 e^+$
arising through gluino dressing diagrams are significant and
comparable to the neutrino modes in large tan$\beta$ models.  Moreover,
it is found that in certain classes of models the Higgsino-mediated
proton decay amplitudes are proportional to a model-dependent
group-theoretical factor which in some cases can be quite small.
There is thus a natural suppression mechanism which can explain without
adjustment of parameters why in the context of SUSY GUTs proton
decay has not yet been seen.  The most interesting such class
consists of $SO(10)$ models in which the dominant flavor-symmetric
contribution to the up-quark mass matrix comes from an effective
operator of the form ${\bf 16}_i {\bf 16}_j {\bf 10}_H {\bf 45}_H$,
where $\langle {\bf 45}_H \rangle$ points approximately in the $I_{3R}$
direction. This class includes a recent model of quark and lepton
masses proposed by the authors.

\end{abstract}
\vspace{-.3cm}
\end{titlepage}

\def\dspace{\baselineskip = .24in}
\dspace

\section{Introduction}

It is well-known that in supersymmetric grand unified theories (SUSY
GUTs) the dominant contribution to proton decay is through dimension-five
operators which arise from the exchange of superheavy, colored
Higgsinos.$^{1,2,3}$ In this paper we shall take a closer look at these
decays.  It will be shown that two
simplifications that are usually introduced in the analysis
of this effect are in certain interesting cases not justified, and
that in going beyond these simplifications some important
features emerge.

The first simplification usually made is to analyze Higgsino-mediated
proton decay in the context of the minimal SUSY $SU(5)$ model.
In that model there is a very simple relation between the
proton-decay amplitudes and the mass matrices of the quarks and
leptons that allows one, in fact, to write these amplitudes
in terms of the quark masses and the Kobayashi-Maskawa matrix elements.
This relation comes from the fact that the Yukawa couplings of
the colored Higgsinos which mediate proton decay are equal, because
of $SU(5)$ invariance, to
the Yukawa couplings of the ordinary light Higgs doublets, $H$ and
$H'$, that give rise to light fermion masses. But this cannot be
taken seriously precisely because the minimal SUSY $SU(5)$ model
gives a wrong account of the light fermion masses. In particular,
minimal $SU(5)$ predicts that $m_e^0 = m_d^0$ and $m_{\mu}^0
= m_s^0$ (superscript zeros refer to quantities evaluated at
the GUT scale), and thus $m_e/m_\mu = m_d/m_s$, which is off by an order
of magnitude.

The second simplification is to neglect gluino loops in dressing
the effective, dimension-five, $\Delta B \neq 0$ operators to
make four-fermion operators. This implies that
only W-ino loops need be considered, which in turn leads to
the conclusion that the only significant modes are those that
involve neutral leptons ($p \rightarrow K^+ \overline{\nu}$ and
$p \rightarrow \pi^+ \overline{\nu}$). It will be explained why
for models with large $\tan \beta$ the gluino loop diagrams
cannot be neglected and decay modes with charged leptons
($p \rightarrow {K}^0 \mu^+$, $p \rightarrow K^0 e^+$) can become
comparable to the neutrino modes.$^4$ (Since the Higgsino-mediated
proton-decay rate
goes as $\tan^2 \beta$ it might be thought that the case of
large $\tan \beta$ is in conflict with present limits.  While this may
be true in minimal SUSY $SU(5)$ model,
there are several likely suppression mechanisms which have
been proposed in the literature$^5$ in going beyond minimal $SU(5)$
and one such group theoretical suppression mechanism is
found in section 4 of this paper.  Large $\tan \beta$ is actually
a natural feature of many grand unified models, especially those
based on $SO(10)$ or larger groups where it often is predicted to
be $m_t/m_b$.)

Realistic models of quark and lepton masses in the context of SUSY
GUTs require that the ``bad" $SU(5)$ or $SO(10)$ mass relations
be broken. This means that the quark and lepton masses receive
contributions from additional
operators not present in the minimal $SU(5)$ model, operators
which either involve Higgs fields in representations larger than
the fundamental or are of higher dimension. What this implies is that
the proton-decay amplitudes are related in a less direct,
model-dependent way to the quark and lepton mass matrices. This allows
one in principle to distinguish different theoretical models of
quark and lepton masses by their proton-decay branching ratios, as
will be seen. It will also be seen that an interesting group-theoretical
mechanism exists by which the higgsino-mediated proton decay rate
may be suppressed to levels near but consistent with present limits
without any special adjustment of parameters.
Interestingly, the group structure required for this suppression
is precisely that suggested in a recent model of quark and lepton
masses.$^6$

\section{Review of p Decay in Minimal SU(5)}

\noindent
{\large\bf (a) W-ino loops:}

\vspace{0.5cm}

\noindent
In order to set the stage for the later analysis it is
convenient to review briefly the standard analysis of
Higgsino-mediated proton decay in the minimal SUSY $SU(5)$
model. In that model the Yukawa couplings of the quarks
and leptons come from the following terms in the superpotential.

\begin{equation}
W_{{\rm Yukawa}} = \frac{1}{2} \sum_{i,j = 1}^{3} U_{ij} \left[ {\bf 10}_i
{\bf 10}_j \right] {\bf 5}_H/v + \sum_{i,j = 1}^{3} D_{ij} \left[
\overline{{\bf 5}}_i {\bf 10}_j \right] \overline{{\bf 5}}_H/v',
\end{equation}

\noindent
where $U_{ij}$ and $D_{ij}$ are the mass matrices of the charge-$\frac{2}{3}$
and charge-$\frac{1}{3}$ quarks.
In terms of multiplets of the standard model gauge group this gives

\begin{equation}
W_{{\rm Yukawa}}(H,H') =  \sum_{ij} U_{ij} \left[ u_i^c  Q_j \right] H/v +
\sum_{ij} D_{ij} \left[ d_i^c Q_j + L_i l_j^+ \right] H'/v',
\end{equation}

\noindent
and
\begin{equation}
W_{{\rm Yukawa}}(H_C,H_C') =  \sum_{ij} U_{ij} \left[ \frac{1}{2} Q_i Q_j
+ u_i^c l_j^+ \right]
H_C/v + \sum_{ij} D_{ij} \left[ L_i Q_j + d_i^c u_j^c \right] H_C'/v'.
\end{equation}

\noindent
{}From the diagrams of Fig. 1 there arise two types of B-violating quartic
terms involving the interactions in eq. 3. These are given by

\begin{equation}
W_{\Delta B \neq 0} = \frac{1}{M_C v v'} \sum_{ij} \sum_{kl}
D_{ij} U_{kl} \left[ \frac{1}{2} (L_i Q_j) (Q_k Q_l) + (d_i^c u_j^c)
(u_k^c l_l^+) \right],
\end{equation}

\noindent
where $M_C$ is the mass of the superheavy color-triplet Higgsino.
Here the contractions of $SU(2)_L$ and $SU(3)_c$ indices are as follows:
\begin{eqnarray}
(L_iQ_j)(Q_kQ_l) &=& \epsilon_{\alpha \beta \gamma}(\nu_i d_j^\alpha-
e_i u_j^\alpha)(u_k^\beta d_l^\gamma-d_k^\beta u_l^\gamma) \nonumber \\
(d_i^cu_j^c)(u_k^cl_l^+) &=& \epsilon_{\alpha \beta \gamma}
(d_i^{c \alpha} u_j^{c \beta}) (u_k^{c \gamma} l_l^+)~. \nonumber
\end{eqnarray}

In terms of component fields the operators in eq. 4 are of dimension
five and contain two fermions and two scalars (squarks and/or sleptons).
Since squarks and sleptons are heavier than the proton they must be
converted into quarks and leptons by a gluino, W-ino or neutralino.
This means that the diagram in Fig. 1 must be dressed by a vertex
loop.

It is generally argued that the gluino loops can be neglected. This
argument can be stated in the following way. The products of superfields
$Q_jQ_kQ_l$ and $d_i^c u_j^c u_k^c$ appearing in eq. 4 are totally
antisymmetric in color. Since they are bosonic, they must also be totally
antisymmetric in flavor. (Note: flavor, not family.) Now suppose
that one neglects all flavor-dependence introduced by the gluino loop.
Then, in the resulting four-fermion operators, the products of quark
fields, $\psi_{Q_j} \psi_{Q_k} \psi_{Q_l}$ and
$\psi_{d_i^c} \psi_{u_j^c} \psi_{u_k^c}$, will have the original
flavor structure of eq. 4, namely they will be totally antisymmetric
in flavor. But since these are now products of fermions
their antisymmetry in color and flavor implies that they must
also be totally antisymmetric in spin, which is clearly impossible
for three spin-$\frac{1}{2}$ objects.

The gluino loops, therefore, are only important to the extent that
they introduce some additional flavor dependence. This can happen
through the nondegeneracy of the different flavors of squark.
But that gives a GIM-like suppression factor which is known to be very
small, especially for the squarks of the first two families,
because of the limits on squark non-degeneracy from the neutral Kaon
system.  Mixing in the squark/slepton mass--squared matrix
is another source of flavor dependence.
(In addition, there
is a suppression factor of $m_u$ that comes from the fact that the four-fermion
operator must involve only $u$ rather than $c$ or $t$ quarks.  But this
is not an extra suppression, because including flavor mixing in the
squark masses, this $m_u$ factor will be replaced by $m_c$ or $m_t$.)
What shall be
seen in section 2(b), however, is that if $\tan \beta$ is large
flavor-change arises in the squark mass matrix which, though
very small, is nevertheless enough to make the gluino-loop
contributions to proton decay significant and comparable to the
W-ino contribution.

Proceeding with the standard analysis, one concludes that the dominant
contribution to proton decay comes from dressing the effective,
dimension-five operators coming from eq. 4 with W-ino loops.
(The photino and zino loops suffer the same problems as the gluino loops,
with an additional suppression of $\alpha/\alpha_s$. The Higgsino
loops have suppressions of small Yukawa couplings.)
Because the second operator in the brackets in eq. 4 contains
only $SU(2)_L$-singlet fields it can therefore be neglected.

{}From the form of the first term in eq. 4 it is readily seen that
the W-ino loop leads to the following effective four-fermion
operators:

\begin{equation}
\begin{array}{ccl}
{\cal L} & = & \frac{1}{M_C v v'} \frac{\alpha_2}{4 \pi} D_{ji}
U_{kl} \epsilon_{\alpha \beta \gamma} \\
& & \\
& & \times [(u_k^{\alpha} d_l^{\beta})(d_i^{\gamma} \nu_j)(f(d_k,u_l)
+ f(d_k,u_l)) \\
& & \\
& & + (d_k^{\alpha} u_l^{\beta})(u_i^{\gamma} l_j^-) (f(u_k,d_l)
+ f(d_i,\nu_j)) \\
& & \\
& & + (d_k^{\alpha} u_i^{\beta})(d_l^{\gamma} \nu_j) (f(u_k,d_i)
+ f(u_l,l_j^-)) \\
& & \\
& & + (u_k^{\alpha} d_i^{\beta})(u_l^{\gamma} l_j^-) (f(d_k,u_i)
+ f(d_l,\nu_j)) ].
\end{array}
\end{equation}

\noindent
Here the color indices ($\alpha$, $\beta$, $\gamma$) and
flavor indices ($i$, $j$, $k$, $l$) are understood to be summed
over, and the fermion fields paired together in parentheses are
spin-contracted to singlets.
$f$ is a loop integral with dimension of $M^{-1}$ defined by
$f(a,b) \equiv \frac{m_{\tilde{W}}}{m_{\tilde{a}}^2 - m_{\tilde{b}}^2}
\left( \frac{m_{\tilde{a}}^2}{m_{\tilde{a}}^2 - m_{\tilde{W}}^2}
\ln \frac{m_{\tilde{a}}^2}{m_{\tilde{W}}^2} - [a \rightarrow b] \right)$.
For $m_{\tilde{a}} \simeq m_{\tilde{b}} \gg m_{\tilde{W}}$,
$f(a,b) \simeq \frac{m_{\tilde{W}}}{m_{\tilde{a}}^2}$ and for
$m_{\tilde{a}} \simeq m_{\tilde{b}} \ll m_{\tilde{W}}$,
$f(a,b) \simeq {1 \over {m_{\tilde{W}}}} {\rm ln} {m_{\tilde{W}}^2
\over {m_{\tilde{a}}^2}}$.  While deriving eq. (5), it has been assumed that
flavor mixing in the squark sector is negligible, which should hold to
a good approximation in supergravity models.  We have verified that
the effect of flavor mixing that occurs in large tan$\beta$ models
(see discussion in section 2.(b)) is negligible in so far as the
W-ino dressing graphs are concerned.

If one neglects contributions proportional to $m_u$, then it is
easily seen that only the third term in the brackets in eq. 5
contributes to proton
decay. One can rewrite that term in the physical basis of the
charge-$\frac{2}{3}$ quarks, where $U_{ij}$ is diagonal, as follows

\begin{equation}
{\cal L} \propto
\sum_{j = 1}^{3} \sum_{k = 2}^{3} D_{j1} U_{kk}
\epsilon_{\alpha \beta \gamma} (d_k^{\prime \alpha} u_1^{\beta})
(d_k^{\prime \gamma} \nu_j),
\end{equation}

\noindent
where we have taken the $f$ factors to be flavor independent
(which should be a very good approximation since the squarks
(and sleptons) must be nearly degenerate.)
Here $d_i' = \sum_{j} (V_{KM})_{ij} d_j$.
The important conclusion is immediate that only the
neutrino modes of proton decay are important.

Note the (paradoxical) result that the ratio of $d$ to $s$ quarks,
which determines the branching ratio of $\pi$ to $K$ modes,
is controlled by the elements of the matrix $U_{ij}$. In particular,
the ratio of the coefficients
of the $(d u)(d \overline{\nu})$ and $[(s u)(d \overline{\nu}) +
(d u)(s \overline{\nu})]$ terms is (neglecting terms proportional
to $m_u$) given by

\begin{equation}
\frac{C_{dd}}{C_{ds}} \cong
\frac{m_c \sin^2 \theta_c + A_s m_t V_{td}^2}
{- m_c \sin \theta_c - A_s m_t V_{td} V_{cb}}.
\end{equation}

\noindent
Or:

\begin{equation}
\frac{C_{dd}}{C_{ds}} \cong - \sin \theta_c \left(
\frac{1 + a^2 b}{1 - ab} \right),
\end{equation}

\noindent
where $a \equiv \frac{V_{td}}{V_{ts} \sin \theta_c}$, and $b \equiv
\frac{V_{ts}^2 m_t A_s}{m_c}$ are quantities that can be measured in
low-energy experiments. (Except for the phase of $b$. $m_c/m_t$
has in general a non-trivial phase which cannot be measured
in low energy experiments, but which can be predicted in specific
models of quark and lepton masses.)  $A_s$ is a short distance
renormalization factor, proportional to the Yukawa couplings.  If
tan$\beta$ is small (tan$\beta \le 10$), $A_s = (1-Y_t/Y_f)^{1/12}$
where $Y_t = h_t^2$ is the top--quark Yukawa coupling--squared and
$Y_f$ is its `true' fixed point values, i.e., the value of $Y_t$
at the weak scale if
$Y_t$ were infinite at the GUT scale. ($Y_f \simeq 1.29$ corresponding
to $\alpha_s(M_Z) = 0.12$).  For tan$\beta = m_t/m_b$, $A_s = 1$.  If
tan$\beta$ takes intermediate values, $A_s$ is to be evaluated
numerically.  The parameters $a$ and $b$
are of order unity, present experiments allow a factor of 2 uncertainty
in their numerical values.

\vspace{1cm}

\noindent
{\large\bf (b) Gluino loops, large $\tan \beta$ and charged lepton modes:}

\vspace{0.5cm}

\noindent
The coefficient of the effective dimension-five operator,
$L_iQ_jQ_kQ_l$, in eq. 4
depends on the mass matrix of the charge-$\frac{2}{3}$ quarks, $U_{ij}$.
Thus, a gluino loop which dressed this operator would be
correspondingly enhanced by containing a virtual $t$ squark.
Such a diagram is shown in Fig. 2. Since, however, the external
charge-$\frac{2}{3}$ quarks must be $u$'s, for this to happen
a flavor change must occur in the gluino loop, such as is indicated by the
$\tilde{t}^* \tilde{u}$ squark mass insertion shown in Fig. 2.
This gives a compensating suppression; the question is by how much.
If large enough flavor change can occur in the squark mass matrix,
then the gluino loops can be important. As will now be explained, this
is indeed the case for large $\tan \beta$.$^4$

If one neglects the Yukawa couplings of the lighter two generations
then there are only two Yukawa terms left. These can be written

\begin{equation}
{\cal L}_{{\rm Yukawa}} = \frac{m_b}{v'} b^c \left(
\begin{array}{c} t' \\ b \end{array} \right)
\times \left( \begin{array}{c} H^{\prime 0} \\ H^{\prime -}
\end{array} \right) + \frac{m_t}{v} t^c \left(
\begin{array}{c} t \\ b' \end{array} \right) \times
\left( \begin{array}{c} H^+ \\ H^0 \end{array} \right).
\end{equation}

\noindent
Here and throughout, the unprimed quark fields denote the
mass eigenstates, while $t'$ denotes the $SU(2)_L$ partner
of $b_L$, and $b'$ denotes the $SU(2)_L$ partner of $t_L$.
Thus $t' = V_{tb}^* t + V_{cb}^* c + V_{ub}^* u$ and
$b' = V_{tb} b + V_{ts} s + V_{td} d$.

Corresponding to eq. 9 there are soft SUSY-breaking trilinear
terms

\begin{equation}
{\cal L}_{{\rm soft}} = A \frac{m_b}{v'} \tilde{b}^c
\left( \begin{array}{c} \tilde{t}' \\ \tilde{b} \end{array}
\right) \times \left( \begin{array}{c} H^{\prime 0} \\
H^{\prime -} \end{array} \right) + A \frac{m_t}{v}
\tilde{t}^c \left( \begin{array}{c} \tilde{t} \\
\tilde{b}' \end{array} \right) \times \left(
\begin{array}{c} H^+ \\ H^0 \end{array} \right).
\end{equation}

\noindent
Here and throughout, the squark fields $\tilde{t}$,
$\tilde{b}$, $\tilde{c}$ etc. are not the squark-mass
eigenstates but the superpartners of the quark-mass
eigenstates $t$, $b$, $c$ etc. Similarly, $\tilde{t}'$
and $\tilde{b}'$ are the superpartners of $t'$ and
$b'$.

The interactions given in eq. 10 give rise to the diagrams
shown in Fig. 3. There are also related diagrams in which
the vertices are the hard, SUSY-invariant Yukawa couplings
and the SUSY breaking comes from the non-equality of the squark and
quark masses. Altogether, these give effective low-energy terms
of the form

\begin{equation}
\Delta {\cal L} \approx
m_0^2 (3+A^2) (16 \pi^2)^{-1} \ln
\left( \frac{M_{{\rm GUT}}}{M_{{\rm SUSY}}} \right)^2
\left[ c_1 \left( \frac{m_b}{v'} \right)^2 \tilde{t}^{\prime *} \tilde{t}'
+ c_2 \left( \frac{m_t}{v} \right)^2 \tilde{b}^{\prime *} \tilde{b}'\right],
\end{equation}

\noindent
where $c_1$ and $c_2$ are numbers of order one that depend on the
precise spectrum of sparticles.  eq. (11) with $c_1=c_2=1$
is what one would obtain with supergravity boundary conditions
at the GUT scale for the squark flavor mixing, if one ignores
the running of all the relevant parameters.  This is, of course, very
naive, but this should still tell us the correct results to within a
factor of 3 or so.  $c_1$ and $c_2$ parametrizes
the effect of such running.
Substituting
$\tilde{t}' = V_{tb}^* \tilde{t} + V_{cb}^* \tilde{c}
+V_{ub}^* \tilde{u}$, one sees that, for example, a $\tilde{u}$-$\tilde{t}$
mixing of order $\left( \frac{m_b}{v'} \right)^2 V_{ub}$ is produced.
For large tan$\beta$, ${m_b \over v'} \sim 1$ and such flavor mixing
becomes significant.

Armed with this result one can go back and evaluate the
graph shown in Fig. 2. One obtains (in the minimal $SU(5)$
model)

\begin{equation}
{\cal L}_{{\rm eff}}^{{\rm gluino}} \sim \frac{\alpha_s}{4 \pi}
\left( \frac{m_b}{v'} \right)^2
\frac{1}{M_C v v'} \sum_{ij} D_{i1} U_{33} V_{ub}^* V_{tj}
\left[ (l_i^- u - \nu_i d') (d_j u) \right].
\end{equation}

\noindent
This is to be compared to the dominant W-ino graph contribution
which gives

\begin{equation}
{\cal L}_{{\rm eff}}^{{\rm Wino}} \sim \frac{\alpha_2}{4 \pi}
\frac{1}{M_C v v'} \sum_i D_{i1} U_{22} \left[
(s' u) (s' \nu_i) \right].
\end{equation}

\noindent
Since $U_{33}/U_{22} = m_t^0/m_c^0$ one sees that for $m_b/v'
\sim 1$ (that is, for $\tan \beta \sim m_t/m_b$)
the gluino and W-ino contributions are comparable,
and therefore the charged lepton and neutrino modes are
comparable.

To be more precise, we shall present below the effective Lagrangian
arising through gluino dressing to lowest order in the flavor mixing
parameters $\Delta_{ij}^{u,d} \equiv (m_{u,d}^2)_{ij}/[(m_{u,d}^2)_{ii}
-(m_{u,d}^2)_{jj}]$ for $i \ne j$ for the case of minimal SUSY $SU(5)$:
\begin{eqnarray}
{\cal L}_{eff}^{\rm gluino}  &=&  -{4 \over 3} {\alpha_s \over 4 \pi}U_i
V_{kl}^* D_l \times
[(u_j^\alpha d_i^{\prime\beta})(u_k^\gamma e_l)\Delta^u_{ij}
\left(f(u_i,d_i')-f(u_j,d_i')\right) \nonumber \\
&+& (u_i^\alpha d_j^{\prime\beta}(u_k^\gamma
e_l)\Delta^d_{ij}\left(f(u_i,d_i')-
f(u_i,d_j')\right) \nonumber \\
&+& (d_i^{\prime\alpha}u_j^\beta)(u_i^\gamma e_l)\Delta^u_{kj}\left(f(u_k,d_i')
-f(u_j,d_i')\right)  \nonumber \\
&+&(d_j^{\prime\alpha} u_k^\beta)(u_i^\gamma e_l)\Delta^d_{ij}\left(f(u_k,d_i')
-f(u_k,d_j')\right) \nonumber \\
 &-&   (u_j^\alpha u_k^\beta)(d_i^{\prime\gamma}e_l)\Delta^u_{ij}\left(
f(u_i,u_k)-f(u_j,u_k)\right) \nonumber \\
&-& (u_i^\alpha u_j^\beta)(d^{\prime\gamma}_ie_l)\Delta^u_{ij}\left(
f(u_i,u_k)-f(u_i,u_j)\right) \nonumber \\
 &-& (u_j^\alpha d_i^{\prime\beta})(d_k^{\prime\gamma}\nu_l)\Delta^u_{ij}\left(
f(u_i,d_i') - f(u_j,d_i')\right) \nonumber \\
&-&(u_i^\alpha d_j^{\prime\beta})(d_k^{\prime\gamma}\nu_l) \Delta^d_{ij} \left(
f(u_i,d_i')-f(u_i.d_j')\right) \nonumber \\
&+& (d_j^{\prime\alpha} d_i^{\prime\beta})(u_i^\gamma \nu_l)\Delta^d_{kj}\left(
f(d_k',d_i')-f(d_j',d_i')\right) \nonumber \\
&+& (d_k^{\prime\alpha}d_j^{\prime\beta})(u_i^\gamma \nu_l)\Delta^d_{ij}\left(
f(d_k',d_i')-f(d_k',d_j')\right) \nonumber \\
&+& (u_j^\alpha d_k^{\prime\beta})(d_i^{\prime\gamma} \nu_l)\Delta^u_{ij}\left(
f(u_i,d_k')-f(u_j,d_k')\right) \nonumber \\
&+& (u_i^\alpha d_j^{\prime\beta})(d_i^{\prime\gamma}\nu_l) \Delta^d_{ij}\left(
f(u_i,d_k')-f(u_i,d_j')\right)]~.
\end{eqnarray}
In the above, we have used a basis where the up--quark mass matrix is
diagonal and the down-quark mass matrix is given by $V_{KM}D$, $D$
being the diagonal down mass matrix.  It should be noted that in the
approximation of keeping only the third family Yukawa couplings, the
mixing parameter $\Delta^d_{ij}$ in this basis is nonvanishing
only for $i=j=3$ because the only mixing is
the down squark sector is of the form $\tilde{b}^{\prime *}\tilde{b'}$.
It is clear from eq. (14) that the dominant
contribution for proton decay via gluino exchange arises from the
first and the fifth terms, all other terms being proportional to the
small up--quark mass.  eq. (14) justifies the qualitative discussion
preceeding it.  In particular, if the SUSY spectrum is
such that the gluino is lighter than the squarks, since $m_{\tilde{g}}
\simeq (\alpha_s/\alpha_2) m_{\tilde{W}}$, the gluino dressing effect
has a factor of $4/3 (\alpha_s/\alpha_2)^2\sim 18$ in the amplitude
relative to W-ino dressing.
Using the naive estimate of $c_1=1$ in eq. (11) and $A=1$, we find that
the two amplitudes are about the same.

It should be noted that, just as for the W-ino loops,
the second operator in eq. 4, namely the $d^c u^c u^c l^+$
operator, can be neglected in the gluino loops. This is because
there is no flavor-changing in the right-handed
squark masses coming from diagrams analogous to Fig. 3.
If one interchanges $\tilde{t}'$ and $\tilde{b}^c$
in Fig. 3a, for example, one merely gets the effective
flavor-conserving operator $\tilde{b}^{c *} \tilde{b}^c$.

\section{Models With Realistic Fermion Masses}

So far we have been considering the unrealistic
case where the quark and lepton Yukawa couplings are those
of minimal SUSY $SU(5)$. As we have seen, what
controls proton decay in the general case (for both W-ino and
gluino loops) is the Yukawa coupling of the colored
Higgsino, $H_C$, to the quark bilinear $Q_i Q_j$ and of the
$H_C'$ to $L_i Q_j$. For minimal
SUSY $SU(5)$ the former is given by the up quark mass matrix, $U_{ij}$,
as shown in eq. 3, but in general it will be some different matrix,
which we will denote $U_{ij}'$: $W_{{\rm Yukawa}}(H_C)
= \sum_{ij} U_{ij}' \left[ Q_i Q_j \right] H_C/v + ...$. Similarly,
the coupling of $H_C'$ is given by some matrix $D'_{ij}$ which is
not in general the same as $D_{ij}$.
Let us now examine the relationship
between $U_{ij}'$ and $U_{ij}$ in some simple cases.

The matrix $U_{ij}$ can be written as the sum of a symmetric and an
antisymmetric piece:

\begin{equation}
U_{ij} =  f_{(ij)} + g_{[ij]}.
\end{equation}

\noindent
Let us assume for simplicity that {\it there is only a single term
that contributes to the symmetric piece}. Then the matrix $U_{ij}'$
will have the form

\begin{equation}
U_{ij}' = r \cdot f_{(ij)}.
\end{equation}

\noindent
It is easily seen that the antisymmetric piece $g_{[ij]}$ does not
contribute to $U_{ij}'$, for $\epsilon_{\alpha \beta \gamma} Q_i^{\alpha}
Q_j^{\beta} H_C^{\gamma}$ being antisymmetric under $SU(2)_L$ and
$SU(3)_c$ has to be symmetric in flavor. The factor $r$ is
group-theoretical and gives the ratio of the coupling of the
color-triplet Higgs(ino) to the doublet Higgs(ino).
This factor will be of great interest to us later.

To see the branching ratios in proton decay we want to go to the
physical basis of the charge-$\frac{2}{3}$ quarks, as in the discussion
leading to eq. 6.
This is done by some transformation under which $Q_i \rightarrow
W_{ij} Q_j$,
$u^c_i \rightarrow \tilde{W}_{ij} u^c_j$, and $U \rightarrow
\tilde{W}^T U W$, so that
$U'$ gets transformed as $U' \rightarrow W^T U' W$. It is
interesting to see what happens if there is
a hierarchy among the elements of $U$, as is usually the case
in models of quark and lepton masses. Assume, therefore that $f_{33}
> f_{23} \sim g_{23} > f_{22}$, and that the elements of the first row and
column can be neglected. Then one has, approximately, that

\begin{equation}
U \rightarrow \left( \begin{array}{ccc} 0 & 0 & 0 \\
0 & f_{22} - \frac{f_{23}^2 - g_{23}^2}{f_{33}} & 0 \\
0 & 0 & f_{33} \end{array} \right) = \left( \begin{array}{ccc}
0 & 0 & 0 \\ 0 & m_c & 0 \\ 0 & 0 & m_t \end{array} \right),
\end{equation}

\noindent
while

\begin{equation}
U' \rightarrow r \cdot \left( \begin{array}{ccc} 0 & 0 & 0 \\
0 & f_{22} - \frac{f_{23}^2 - g_{23}^2}{f_{33}} & g_{23} \\
0 & g_{23} & f_{33} \end{array} \right) =
r \cdot \left( \begin{array}{ccc} 0 & 0 & 0 \\
0 & m_c & g_{23} \\
0 & g_{23} & m_t \end{array} \right).
\end{equation}

\noindent
One sees that the {\it relative} contributions to proton decay from the (2,2)
and (3,3) elements are the same as in the minimal
SUSY $SU(5)$ model.
But now there is also the contribution from the (2,3) element,
which is obviously model-dependent. In particular models
of quark and lepton masses $g_{23}$ would in general be
known. If $\tan \beta \ll m_t/m_b$ then the W-ino loops dominate
and proton decay is controlled by the terms in eq. 6 with
the matrices $D$ and $U$ replaced by $D'$ and $U'$. One can then
generalize eqs. 7 and 8 using eq. 18.

\begin{equation}
\frac{C_{dd}}{C_{ds}} \cong - \sin \theta_c
\left( \frac{1 + a^2 b - 2 a b^{\frac{1}{2}} \overline{g}_{23}}
{1 - a b - (a - 1) b^{\frac{1}{2}} \overline{g}_{23}} \right),
\end{equation}

\noindent
where $\overline{g}_{23} \equiv g_{23}/(m_c m_t)^{\frac{1}{2}}$
has been normalized so that it is typically a number of order unity (if
non--zero).  Thus

\begin{equation}
\frac{\Gamma(p \rightarrow \pi \overline{\nu})/\Gamma(p \rightarrow
K \overline{\nu})}{[\Gamma(p \rightarrow \pi \overline{\nu})
/\Gamma(p \rightarrow K \overline{\nu})]_{{\rm min~SU(5)}}}
=\left| \frac{1 - 2 \left( \frac{a b^{\frac{1}{2}}}{1 + a b^2} \right)
\overline{g}_{23}}{1 - \frac{(a-1)b^{\frac{1}{2}}}{1 - a b}
\overline{g}_{23}} \right|^2.
\end{equation}

\noindent
One can straightforwardly derive the analogous expression
when the first row and column of $U'$ cannot be neglected.

What one sees is that by measuring the $\pi^+ \overline{\nu}$ to
$K^+ \overline{\nu}$
ratio in proton decay one could distinguish different
models of quark and lepton masses, since in different
models $\overline{g}_{23}$ would take different ---
and for predictive models, computable --- values.

If $\tan \beta$ is large, then these neutrino modes can get
comparable contributions from both the W-ino and
gluino loops. This would make the expression for
the branching ratio $\Gamma (p \rightarrow \pi^+ \overline{\nu})/
\Gamma (p \rightarrow K^+ \overline{\nu})$ much more
complicated (unless $\tan \beta$ is large and the spectrum is such
that the gluino
loop dominated). One can tell the relative importance
of the W-ino loops and gluino loops by the relative
importance of the neutrino and charged lepton modes, of
course.

A further test of models comes, if $\tan \beta$ is large,
from examining the ratio of $e^+$ to $\mu^+$. This probes
the value of $D'_{11}/D'_{21}$, as can be seen from
eq. 12, where $D$ and $U$ should be replaced in the general case
by $D'$ and
$U'$. In minimal $SU(5)$, $D'_{ij} = D_{ij}$ and
$\Gamma(p \rightarrow {K}^0 e^+)/\Gamma(p \rightarrow
{K}^0 \mu^+) \cong \left| D_{11}/D_{21} \right|^2
\cong \left| m_d/m_s \sin \theta_c \right|^2 \cong 0.05$.
(Recall that we are working in the basis where $U_{ij}$ is
diagonal, so that $D_{ij}$ is nondiagonal.) In a realistic model
of quark and lepton masses this branching ratio would be different.
For example, in the Georgi-Jarlskog model$^7$ it comes out to be
about 0.02.

Finally, for large $\tan \beta$, the branching ratios
$\Gamma(p \rightarrow {K}^0 l^+)/\Gamma(p
\rightarrow \pi^0 l^+)$ give independent information
about the elements of $U'_{ij}$. (Cf. eq. 12.)
\section{Group-theoretical Suppression of p Decay}

Where eq. 16 applies one sees that the proton decay
amplitudes coming from either W-ino or gluino loops
are proportional to the group-theoretical factor $r$.
This factor is given for various simple cases in
Table I.

In discussing the value of $r$ it should be kept in mind that
the form given in eq. 16, in which $r$ appears, is only valid
for cases in which the symmetric contribution to $U_{ij}$
comes from a single term (or at least from several
terms which all have the same gauge structure).
The simplest possibility in $SU(5)$ is that the symmetric
contribution to $U_{ij}$ comes from a ${\bf 5}$ of Higgs,
as in the minimal $SU(5)$ model.
In that case the couplings of doublet and triplet Higgs
are trivially the same and $r = 1$. The other $SU(5)$ multiplet
that can couple flavor-symmetrically to the charge-$\frac{2}{3}$
quarks is the ${\bf 50}_H$. But, while this contains a color-triplet
it does not contain the usual type of Higgs doublet, so that
if {\it only} a ${\bf 50}_H$ were present then $U_{ij}$ would
vanish and $r$ would be infinite, which is clearly unrealistic.
If both a ${\bf 5}_H$
and a ${\bf 50}_H$ contribute then eq. 16 does not apply.

One can also consider
higher order operators involving bilinears of Higgs fields.
Suppose that the flavor-symmetric contribution to $U_{ij}$
comes from the operator $\left[ {\bf 10}_i {\bf 10}_j \right]
{\bf 5}_H {\bf 24}_H$.
The product ${\bf 5} \times {\bf 24}$ contains ${\bf 5} + {\bf 45}
+ {\bf 70}$. Of these, ${\bf 70}$ does not couple to ${\bf 10}
\times {\bf 10}$, and ${\bf 45}$ couples antisymmetrically.
Thus there is only one term that couples symmetrically in flavor
and eq. 16 does apply with a value $r = - \frac{2}{3}$, as can
be straightforwardly shown.

The next simplest possibility is that the symmetric
contribution to $U_{ij}$ comes from the bilinear ${\bf 45}_H \times
{\bf 24}_H$. There are two independent ways to contract this to get
a flavor symmetric term, namely into a ${\bf 5}$ or a ${\bf 50}$,
so eq. 16 does not apply unless for some reason only a single
contraction or linear combination of contractions of ${\bf 45}_H
\times {\bf 24}_H$ contributes.
For example, if only the operator $\left[ {\bf 10}_i {\bf 10}_j \right]
({\bf 45}_H {\bf 24}_H)_{{\bf 5}}$ appears, where ${\bf 45}_H \times
{\bf 24}_H$ is contracted in the ${\bf 5}$ channel, then $r = 2/\sqrt{3}$.
(Such a contraction might come about by integrating out a ${\bf 5} +
\overline{{\bf 5}}$, as shown in Fig. 4a. However, in that case
there is also in general the direct ${\bf 10}_i {\bf 10}_j {\bf 5}_H$
term, which, if it contributes significantly to $U_{ij}$ and $U_{ij}'$,
invalidates eq. 16.) A different term arises from integrating out
a ${\bf 10} + \overline{{\bf 10}}$, as shown in Fig. 4b. If this gives the
only symmetric contribution to $U_{ij}$ then $r = 0$. (The reason for this
is that the color triplet in the ${\bf 45}_H$ does not have a coupling to
a pair of quark doublets. See the discussion above of why $g_{[ij]}$ does
not contribute to $U_{ij}'$.) This would be a way of suppressing
Higgsino-mediated proton decay group-theoretically in $SU(5)$.
However, $SU(5)$ seems to require an elaborate Higgs structure involving
${\bf 75}_H + {\bf 50}_H + \overline{{\bf 50}}_H$ to solve the doublet-
triplet-splitting problem, and the mechanism being discussed here
would require in addition
a ${\bf 45}_H$. We thus consider the $SO(10)$ example of a group-theoretical
suppression of proton decay discussed below to be more interesting.

Turning now to $SO(10)$, the simplest possibility is that
the sole symmetric contribution to $U_{ij}$ is from a
${\bf 10}_H$, which, like the case of the fundamental Higgs
in $SU(5)$, gives $r = 1$. On the other hand, if the sole contribution
to $U_{ij}$ is from a $\overline{{\bf 126}}_H$ then $r = \sqrt{3}$.
(It should be noted that there are two color-triplets in the
$\overline{{\bf 126}}_H$, or more precisely two $(3,1,-\frac{1}{3})$
representations of $SU(3)_c \times SU(2)_L \times U(1)_Y$. $\left[
Q_i Q_j \right]$ couples to one
linear combination of these.
$r = \sqrt{3}$ is the ratio of the strength of {\it this}
coupling to the strength of the Higgs doublet's coupling to $\left[
u_i^c Q_j \right]$. But the proton decay amplitude will also
depend on the mixing angle which tells how much of the lightest
color-triplet Higgsino is contained in the linear combination
that couples to $\left[ Q_i Q_j \right]$.)

The simplest Higgs bilinear to be considered in
$SO(10)$ is ${\bf 10}_H \times {\bf 45}_H$.
This contains ${\bf 10} + {\bf 120} + {\bf 320}$. Since the
${\bf 320}$ does not couple to $\left[ {\bf 16}_i {\bf 16}_j \right]$,
and ${\bf 120}$
couples antisymmetrically, there is only one symmetric contribution
from this term and eq. 16 applies. The value of $r$ depends
on what direction in group space $\langle {\bf 45}_H \rangle$ points in.
Let us call
that direction, which is a linear combination of $SO(10)$ generators,
$Q$.
Then it is easily seen that $r = \left( \frac{2 Q_{(u,d)}}{Q_{(u,d)}
+ Q_{u^c}} \right)$. (Note that the same expression
applies to the $SU(5)$ case of ${\bf 5}_H \times {\bf 24}_H$
with $Q = Y/2$ and gives $r = 2(\frac{1}{6})/(\frac{1}{6}
-\frac{2}{3}) = -\frac{2}{3}$.)
This is
made small if  $Q$ points approximately in the $I_{3R}$
direction, since the left-handed quark doublets are singlets under
$SU(2)_R$. One can write $Q$ as a linear combination of
$I_{3R}$ and $Y/2$. (There is a two-dimensional space
of generators of $SO(10)$ that commute with the generators
of $SU(3)_c \times SU(2)_L \times U(1)_Y$.) Since $Y/2$
is non-zero for the left-handed quark doublets, it is obvious
that $r$ will be small {\it only} if $Q$ points approximately in
the $I_{3R}$ direction.

This result is quite interesting, since, in a recent paper,$^6$
a model of quark and lepton masses is proposed by us
in which the matrix $U_{ij}$ arises from an effective term
$\left[ {\bf 16}_i {\bf 16}_j \right] {\bf 10}_H {\bf 45}_H$ and the
$\langle {\bf 45}_H \rangle$ points approximately in the $I_{3R}$ direction.
Indeed, in that model, $Q \sim I_{3R}$ explains {\it three}
relations among known quantities: the Georgi-Jarlskog
relation $m_{\mu}^0/m_s^0 \approx 3$, the smallness of the
second generation masses compared to those of the third generation,
and the smallness of $V_{cb}$. (In the ``long version" of that model$^4$
$Q \sim I_{3R}$ also explains why $m_c^0/m_t^0 \ll m_s^0/m_b^0$.)
What we have found here is that the same group theoretical
assumption also gives rise to a (needed)
suppression of Higgsino-mediated proton decay.

In the model of Ref. 6 the group-theoretical suppression of the
proton decay
amplitude is numerically of order $10^{-1}$.  This means that, while
suppressed, $p$--decay is still within the reach of Super Kamiokande.

It is worth noting that the term $\left[ {\bf 16}_i {\bf 16}_j
\right] {\bf 10}_H {\bf 45}_H$ has another beautiful property,
which is exploited in Ref. 6. For $i = j = 3$, ${\bf 10}_H \times
{\bf 45}_H$ must be in the symmetric product, and, in particular,
be contracted to into a ${\bf 10}$, which gives equal contributions
to $D_{33}$ and $L_{33}$. But for $i,j = 2,3$ or $3,2$ contractions
into both ${\bf 10}$ and ${\bf 120}$ are allowed, and the latter
contributes differently to $D$ and $L$. Thus a term of the
form $\left[ {\bf 16}_i {\bf 16}_j \right] {\bf 10}_H {\bf 45}_H$
explains why $m_b^0 = m_s^0$ while $m_s^0 \neq m_{\mu}^0$.

\section{Conclusions}

We have shown that in realistic models of quark and lepton masses in
the context of SUSY GUTs, the proton decay rate and branching ratios
are different from the predictions of minimal SUSY $SU(5)$ model.
This would enable one to test quark and lepton mass schemes in SUSY
GUTs by measuring, for example, the branching ratio $B[(p \rightarrow
\pi^+\overline{\nu})/(p \rightarrow K^+\overline{\nu})]$.
In predictive models of fermion masses, such branching ratios are
computable as they are
related to other low energy observables.  We have also emphasized that
in large tan$\beta$ models, as often occurs in $SO(10)$ GUT, the gluino
dressing of the effective $\Delta B \ne 0$ dimension 5 operator is
quite significant.  This opens up the possibility that the charged
lepton decay modes of the proton, $p \rightarrow K^0 \mu^+, p \rightarrow
K^0 e^+$, may be comparable to the neutrino modes.  We give simple
analytic arguments showing how this happens, especially in the case
of tan$\beta = m_t/m_b$ schemes.  The branching ratio $B[(p \rightarrow
K^0 \mu^+)/(p \rightarrow K^0 e^+)]$ will then provide additional clue
to the texture of the fermion mass matrices.  The relative strengths of
the neutral to charged lepton modes will tell us about the parameter
tan$\beta$ itself.  While the precise values of these branching ratios
depend on poorly known (presently) Kobayashi--Maskawa mixing angles
($V_{ub}, V_{td}$), that situation should change in the near future.

We have also found, as discussed in section 4, an interesting group
theoretical suppression mechanism for proton decay rate.  This could
provide a simple answer to the question why in the context of SUSY GUTs,
proton decay has not yet been observed.  We find it interesting that a
recent model$^6$ of quark and lepton masses proposed on quite independent
grounds automatically has this ingredient needed
for group theoretical suppression of proton decay rate.

In closing, let us emphasize that the discovery for proton decay
would not only provide evidence for the violation of baryon number
symmetry near the Planck scale, but it
would also provide, in the context of SUSY GUTs, important clues to
the structure of the fermion mass matrices.  The search for these
decays should continue in earnest.

\newpage

\section*{References}
\begin{enumerate}
\item N. Sakai and T. Yanagida, Nucl. Phys. {\bf B197}, 533 (1982);
S. Weinberg, Phys. Rev. {\bf D26}, 287 (1982).
\item J. Ellis, D.V. Nanopoulos, and S. Rudaz, Nucl. Phys.
{\bf B202}, 43 (1982); P. Nath, A.H. Chamseddine, and R. Arnowitt,
Phys. Rev. {\bf D32}, 2348 (1985); P. Nath and R. Arnowitt,
Phys. Rev. {\bf D38}, 1479 (1988).
\item For a review see J. Hisano, H. Murayama, and T. Yanagida,
Nucl. Phys. {\bf B402}, 46 (1993).
\item T. Goto, T. Nihei and J. Arafune, Preprint ICRR-Report-317-94-12
(1994).
\item G.D. Coughlan, G.G. Ross, R. Holman, P. Ramond, M. Ruiz-Altaba,
and J.W.F. Valle, Phys. Lett. {\bf 158B}, 401 (1985);
J. Hisano, H. Murayama, and T. Yanagida, Phys. Lett. {\bf 291B},
263 (1992); K.S. Babu and S.M. Barr, Phys. Rev. {\bf D48}, 5354
(1993).
\item K.S. Babu and S.M. Barr, Bartol preprint BA-95-11 (hep-ph/9503215).
See also S.M. Barr, Phys. Rev. Lett. {\bf 64}, 353 (1990).
\item H. Georgi and C. Jarlskog, Phys. Lett. {\bf 86B}, 297 (1979).
\end{enumerate}

\newpage

{\large\bf Table I:} The group-theoretical factor $r$ that
enters into the proton-decay amplitude in SUSY GUT models
for which the flavor-symmetric contribution to the mass matrix
of the up quarks comes from a single kind of operator.
The second column gives the product of Higgs fields
appearing in that operator. In the case of the ${\bf 45}_H \times
{\bf 24}_H$ of $SU(5)$ more than one contraction of indices is possible.

\vspace{.2in}

$$\begin{array}{ccc}
{\rm Gauge~group} & {\rm Higgs~operator} & r \\
\hline
SU(5) & {\bf 5}_H & 1 \\
& & \\
SU(5) & {\bf 50}_H & \infty \\
& & \\
SU(5) & {\bf 5}_H \times {\bf 24}_H & -\frac{2}{3} \\
& & \\
SU(5) & ({\bf 45}_H \times {\bf 24}_H)_{{\bf 5}} & 2/\sqrt{3} \\
& & \\
SU(5) & ({\bf 45}_H \times {\bf 24}_H)_{{\rm Fig.2b}} & 0 \\
& & \\
SO(10) & {\bf 10}_H & 1 \\
& & \\
SO(10) & \overline{{\bf 126}}_H & \sqrt{3} \\
& & \\
SO(10) & {\bf 10}_H \times {\bf 45}_H & \frac{2 Q_{(u,d)}}
{Q_{(u,d)} + Q_{u^c}} \\
\end{array}
$$

\newpage

\noindent
{\large\bf Figure Captions}

\vspace{0.3in}

\begin{description}
\item[Fig. 1:] Diagrams involving the exchange
of supermassive, colored Higgsinos which give effective dimension-five
$\Delta B \neq 0$ operators. The dominant contribution
to p decay in general SUSY GUTs comes from (a).
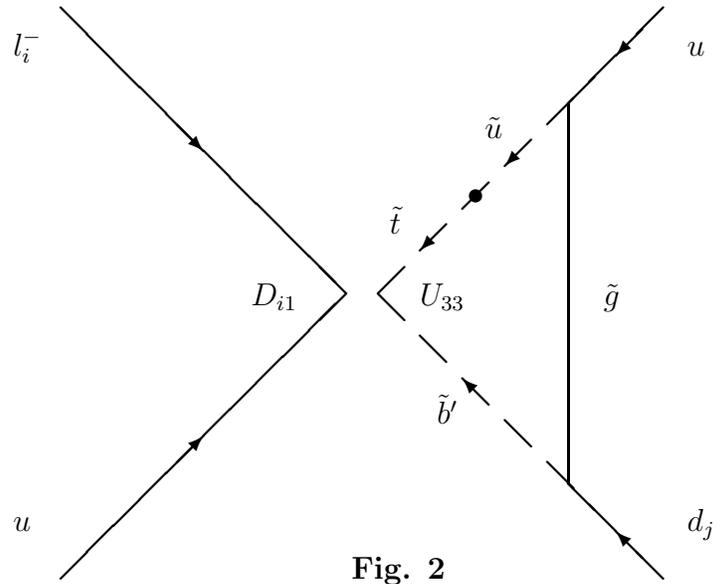
\item[Fig. 2:] A dimension-five $\Delta B \neq 0$ operator
dressed by a gluino loop to produce a four-fermion interaction.
By having a virtual $t$-squark this diagram is enhanced by
a factor of $U_{33} = m_t$. This requires the flavor-changing
squark-mass insertion indicated by the solid circle.
\item[Fig. 3:] A one-loop diagram giving a flavor-changing
squark mass term.
\item[Fig. 4:] Two diagrams that would give operators of the
form ${\bf 10}_i {\bf 10}_j {\bf 45}_H {\bf 24}_H$.
Diagram (a) gives the ${\bf 45}_H$ and ${\bf 24}_H$ contracted
into a ${\bf 5}$. Diagram (b) gives a different contraction,
for which $r=0$.
\end{description}

\newpage

\begin{picture}(360,216)
\thicklines
\put(180,144){\vector(-1,0){47}}
\put(180,144){\vector(1,0){47}}
\put(86,144){\line(1,0){47}}
\put(274,144){\line(-1,0){47}}
\put(175.5,142){$\times$}
\put(36,194){\line(1,-1){50}}
\put(36,94){\line(1,1){50}}
\put(324,194){\line(-1,-1){50}}
\put(324,94){\line(-1,1){50}}
\put(123,153){$H_C'$}
\put(217,153){$H_C$}
\put(11,203){$L_i$}
\put(11,85){$Q_j$}
\put(333,203){$Q_k$}
\put(333,85){$Q_l$}
\put(86,126){$D_{ij}$}
\put(260,126){$U_{kl}$}
\put(162,36){{\bf Fig. 1a}}
\end{picture}

\vspace{0.5in}

\begin{picture}(360,216)
\thicklines
\put(180,144){\vector(-1,0){47}}
\put(180,144){\vector(1,0){47}}
\put(86,144){\line(1,0){47}}
\put(274,144){\line(-1,0){47}}
\put(175.5,142){$\times$}
\put(36,194){\line(1,-1){50}}
\put(36,94){\line(1,1){50}}
\put(324,194){\line(-1,-1){50}}
\put(324,94){\line(-1,1){50}}
\put(123,153){$H_C'$}
\put(217,153){$H_C$}
\put(11,203){$d_i^c$}
\put(11,85){$u_j^c$}
\put(333,203){$u_k^c$}
\put(333,85){$l_l^+$}
\put(86,126){$D_{ij}$}
\put(260,126){$U_{kl}$}
\put(162,36){{\bf Fig. 1b}}
\end{picture}

\newpage

\begin{picture}(360,216)
\thicklines
\put(180,108){\line(1,1){10}}
\put(206,134){\vector(-1,-1){10}}
\put(212,140){\line(1,1){10}}
\put(238,166){\vector(-1,-1){10}}
\put(244,172){\line(1,1){10}}
\put(180,108){\line(1,-1){10}}
\put(196,92){\line(1,-1){10}}
\put(222,66){\vector(-1,1){10}}
\put(228,60){\line(1,-1){10}}
\put(244,44){\line(1,-1){10}}
\put(217,145){\circle*{5}}
\put(252,36){\line(0,1){144}}
\put(288,216){\vector(-1,-1){18}}
\put(252,180){\line(1,1){18}}
\put(288,0){\vector(-1,1){18}}
\put(252,36){\line(1,-1){18}}
\put(60,216){\vector(1,-1){54}}
\put(114,162){\line(1,-1){54}}
\put(60,0){\vector(1,1){54}}
\put(114,54){\line(1,1){54}}
\put(297,198){$u$}
\put(297,18){$d_j$}
\put(42,198){$l_i^-$}
\put(42,18){$u$}
\put(184.5,130.5){$\tilde{t}$}
\put(220.5,166.5){$\tilde{u}$}
\put(202.5,58.5){$\tilde{b}'$}
\put(265.5,103.5){$\tilde{g}$}
\put(132,103.5){$D_{i1}$}
\put(196,103.5){$U_{33}$}
\put(170,0){{\bf Fig. 2}}
\end{picture}

\newpage

\begin{picture}(360,216)
\thicklines
\put(72,72){\line(1,0){20}}
\put(120,72){\vector(-1,0){20}}
\put(128,72){\line(1,0){20}}
\put(156,72){\vector(1,0){20}}
\put(184,72){\line(1,0){20}}
\put(212,72){\line(1,0){20}}
\put(240,72){\line(1,0){20}}
\put(288,72){\vector(-1,0){20}}
\put(128,72){\line(1,2){10}}
\put(145,99){\line(1,1){16}}
\put(170,120){\vector(1,0){20}}
\put(199,115){\line(1,-1){16}}
\put(222,92){\line(1,-2){10}}
\put(174.5,140){$H^{\prime -}$}
\put(87.5,52){$\tilde{t}'$}
\put(174.5,52){$\tilde{b}^c$}
\put(263.5,52){$\tilde{t}'$}
\put(170,0){{\bf Fig. 3a}}
\end{picture}

\vspace{0.5in}

\begin{picture}(360,216)
\thicklines
\put(72,72){\line(1,0){20}}
\put(120,72){\vector(-1,0){20}}
\put(128,72){\line(1,0){20}}
\put(156,72){\vector(1,0){20}}
\put(184,72){\line(1,0){20}}
\put(212,72){\line(1,0){20}}
\put(240,72){\line(1,0){20}}
\put(288,72){\vector(-1,0){20}}
\put(128,72){\line(1,2){10}}
\put(145,99){\line(1,1){16}}
\put(170,120){\vector(1,0){20}}
\put(199,115){\line(1,-1){16}}
\put(222,92){\line(1,-2){10}}
\put(174.5,140){$H^+$}
\put(87.5,52){$\tilde{b}'$}
\put(174.5,52){$\tilde{t}^c$}
\put(263.5,52){$\tilde{b}'$}
\put(170,0){{\bf Fig. 3b}}
\end{picture}

\newpage

\begin{picture}(360,216)
\thicklines
\put(36,174){\vector(1,0){72}}
\put(108,174){\line(1,0){144}}
\put(324,174){\vector(-1,0){72}}
\put(180,138){\vector(0,1){18}}
\put(180,156){\line(0,1){18}}
\put(180,138){\vector(0,-1){18}}
\put(180,120){\line(0,-1){18}}
\put(175.5,136){$\times$}
\put(130,52){\vector(1,1){25}}
\put(155,77){\line(1,1){25}}
\put(230,52){\vector(-1,1){25}}
\put(205,77){\line(-1,1){25}}
\put(225.5,50){$\times$}
\put(95,183){${\bf 10}_i$}
\put(249,183){${\bf 10}_j$}
\put(189,154){${\bf 5}_H$}
\put(189,118){$\overline{{\bf 5}}_H$}
\put(123,36){${\bf 45}_H$}
\put(223,36){${\bf 24}_H$}
\put(162,2){{\bf Fig. 4a}}
\end{picture}

\vspace{0.5in}

\begin{picture}(360,216)
\thicklines
\put(36,144){\vector(1,0){36}}
\put(72,144){\line(1,0){72}}
\put(180,144){\vector(-1,0){36}}
\put(180,144){\vector(1,0){36}}
\put(216,144){\line(1,0){72}}
\put(324,144){\vector(-1,0){36}}
\put(108,96){\vector(0,1){24}}
\put(108,120){\line(0,1){24}}
\put(252,96){\vector(0,1){24}}
\put(252,120){\line(0,1){24}}
\put(175.5,142){$\times$}
\put(247.5,94){$\times$}
\put(95,78){${\bf 45}_H$}
\put(239,78){${\bf 24}_H$}
\put(63,153){${\bf 10}_i$}
\put(139.5,153){${\bf 10}$}
\put(207,153){$\overline{{\bf 10}}$}
\put(279,153){${\bf 10}_j$}
\put(162,42){{\bf Fig. 4b}}
\end{picture}

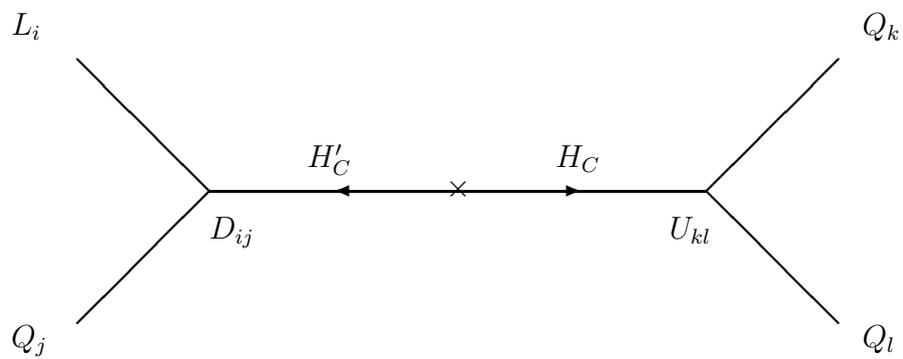
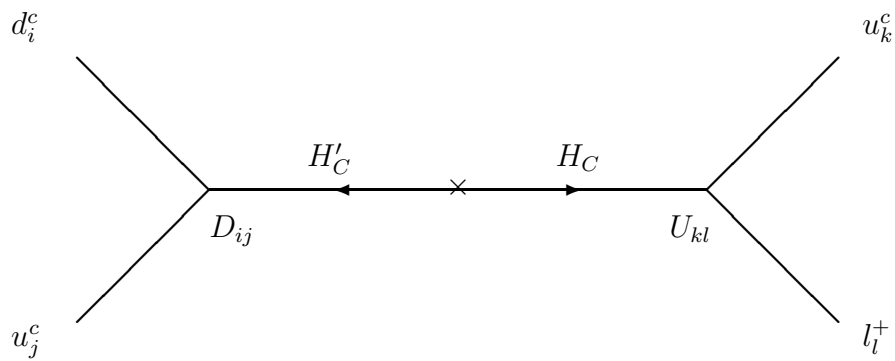
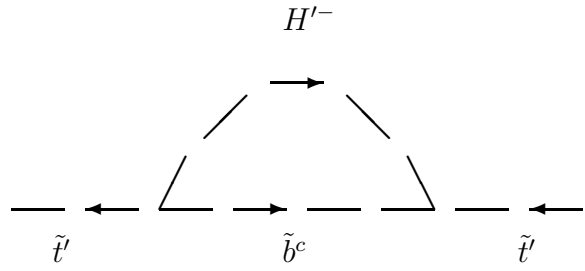
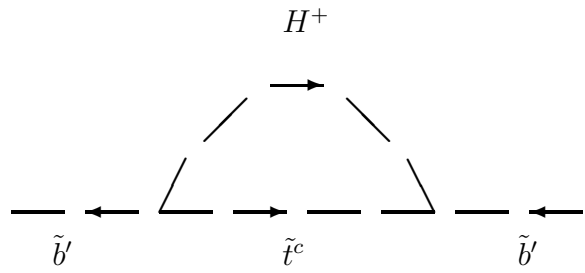
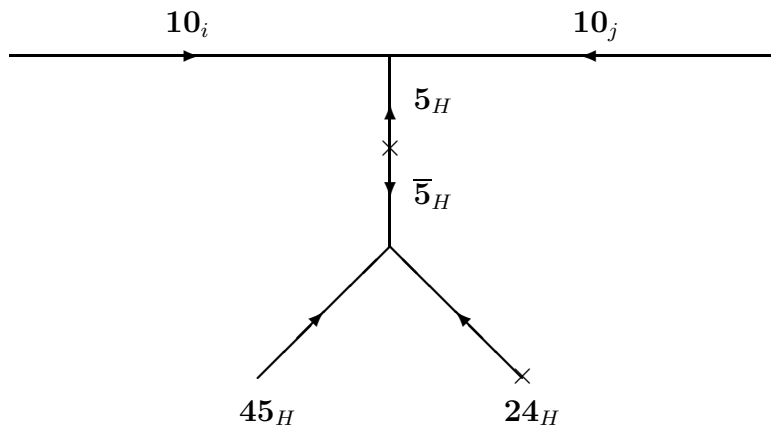
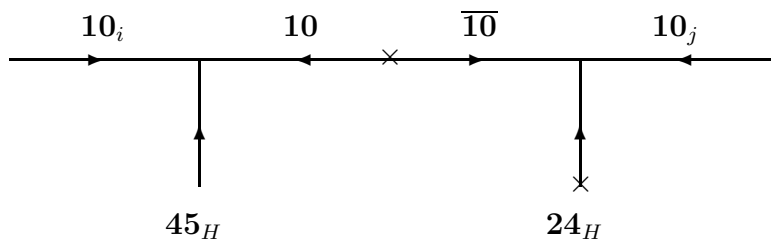
\end{document}